\documentclass[12pt]{spieman}
\usepackage{amsmath,amsfonts,amssymb}
\usepackage{makecell} 
\usepackage{booktabs}
\usepackage{xcolor}
\usepackage{adjustbox}
\usepackage{caption}
\usepackage{url}
\usepackage{subcaption}
\usepackage{graphicx}
\usepackage{placeins}
\usepackage{array}
\usepackage{upgreek}
\usepackage{siunitx}
\usepackage{multirow}
\usepackage{threeparttable}
\usepackage{setspace}
\usepackage{tocloft}
\title{Toward Realistic Numerical Mouse Phantoms to Enable Virtual Imaging-based Assessment of Preclinical Multispectral Dynamic Contrast-Enhanced Photoacoustic Computed Tomography}

\author[a]{Refik Mert Cam}
\author[a]{Hsuan-Kai Huang}
\author[b,c]{Luke Lozenski}
\author[d,e]{Seonyeong Park}
\author[f]{W. Paul Segars}
\author[d,e]{Mark A. Anastasio}
\author[b, g, *]{Umberto Villa}
\affil[a]{University of Illinois Urbana-Champaign, Department of Electrical \& Computer Engineering, 306 North Wright Street, Urbana, IL, United States, 61801}
\affil[b]{The University of Texas at Austin, Oden Institute for Computational Engineering \& Sciences, 201 East 24th Street, Austin, TX, United States, 78712}
\affil[c]{Danmarks Tekniske Universitet, Department of Applied Mathematics and Computer Science, Anker Engelunds Vej 1, Kgs. Lyngby, Denmark, DK-2800}
\affil[d]{Washington University in St$.\,$Louis, Mallinckrodt Institute of Radiology, 510 S Kingshighway Blvd, St. Louis, MO, United States, 63110}
\affil[e]{University of Illinois Urbana-Champaign, Department of Bioengineering, 1406 West Green Street, Urbana, IL, United States, 61801}
\affil[f]{Duke University, Department of Radiology, Box 3808 DUMC
Durham, NC 27710}
\affil[g]{The University of Texas at Austin, Department of Biomedical Engineering, 2107 W Dean Keeton Street, Austin, TX, United States, 78712}

\cftpagenumbersoff{figure}
\cftpagenumbersoff{table} 
\begin{document} 
\maketitle

\begin{abstract}

\textbf{Significance}: Multispectral dynamic contrast-enhanced (DCE) photoacoustic computed tomography (PACT) is a promising modality for characterizing vascular transport processes in preclinical imaging. However, rigorous quantitative validation of multispectral DCE-PACT methods remains challenging because ground truth is unavailable \emph{in vivo}, and physical phantoms offer limited anatomical realism, physiological complexity, and scalability.

\textbf{Aim}: The aim of this study is to develop a computational framework for generating numerical mouse phantoms
for virtual imaging (VI) studies of preclinical multispectral DCE-PACT that enables controlled and reproducible evaluation of reconstruction and concentration-estimation methods by use of physiologically realistic data.

\textbf{Approach}: The framework augments a deformable whole-body mouse anatomical model with vascular structures, heterogeneous tumor morphology, and respiratory and cardiac motion. Contrast-agent kinetics are modeled using the extended Tofts model, and wavelength-dependent optical properties are assigned, with optical absorption incorporating
spatiotemporally varying contrast-agent concentration distributions. A VI case study is conducted to benchmark representative concentration-estimation strategies.

\textbf{Results}: The proposed framework enables controlled and reproducible VI studies of preclinical multispectral DCE-PACT with access to ground-truth contrast-agent distributions. The case study demonstrates 
its utility for benchmarking representative concentration-estimation strategies. 

\textbf{Conclusion}: This work establishes a flexible platform for systematic and physiologically relevant evaluation of preclinical multispectral DCE-PACT methods. More broadly, the proposed framework can support the development and objective assessment of quantitative methods for 
dynamic photoacoustic imaging.

\end{abstract}


{\noindent \footnotesize\textbf{*}Umberto Villa, \linkable{uvilla@austin.utexas.edu} }

\section{Introduction}
Dynamic contrast-enhanced (DCE) imaging
characterizes vascular transport and permeability by tracking the uptake and clearance of an administered contrast agent (CA)~\cite{tofts1999estimating,sourbron2011scope}. Quantitative descriptors of perfusion, permeability, and exchange between vascular and extravascular compartments can be inferred by fitting pharmacokinetic (PK) models to concentration--time curves \cite{tofts1999estimating,sourbron2011scope}. These parameters have been widely employed for tumor characterization, therapy-response monitoring, and assessment of physiological function \cite{tofts1999estimating,sourbron2011scope,barnes2012practical}. In preclinical imaging, DCE approaches are particularly
valuable as they enable controlled investigation of disease progression and therapeutic response in small-animal models~\cite{barnes2012practical}.

Photoacoustic computed tomography (PACT) provides a route to DCE imaging with endogenous hemoglobin absorption contrast and sensitivity to optically absorbing exogenous agents~\cite{wang2012photoacoustic,weber2016contrast}. Rapid tomographic acquisition makes multispectral DCE-PACT well suited for monitoring CA kinetics and biodistribution with high spatiotemporal resolution~\cite{taruttis2012fast,hupple2018light,xiao2018applying}. 
When combined with spectral processing, multispectral measurements can, in principle, provide estimates of time-varying CA
concentrations 
and derived functional biomarkers \cite{taruttis2012fast, xiao2018applying, anani2020quantitative}. This capability is particularly relevant in preclinical small-animal studies, where PACT enables \emph{in vivo} imaging of vascular physiology and molecular processes and supports dynamic assessment of tumor biology, CA kinetics, and treatment response~\cite{xia2013small, laufer2012vivo, xiao2018applying}. 

Despite recent progress, rigorous development and validation of multispectral DCE-PACT methodologies remain challenging~\cite{cox2012quantitative, ding2017constrained, hupple2018light}. Empirical validation using \emph{in vivo} studies and experimental platforms remains essential, but each carries limitations for quantitative assessment~\cite{cox2012quantitative, abadi2020virtual}. In \emph{in vivo} studies, spatially and temporally resolved ground-truth CA 
distributions are typically unavailable; consequently, quantitative estimation error cannot be directly computed, which limits systematic analysis~\cite{abadi2020virtual}. Physical phantoms can provide controlled and repeatable measurement conditions; however, they are commonly simplified and may not faithfully reproduce the coupled physiological, optical, and acoustic effects
encountered in DCE-PACT \cite{else2024effects,durairaj2020unsupervised,rosenthal2009quantitative,vogt2016biologically}. Moreover, construction of large ensembles of physical phantoms that capture
biologically relevant variation in anatomy, vascular architecture, tumor morphology, and motion
can be costly and impractical \cite{kiarashi2015development,keenan2016design}.

Virtual imaging (VI) studies provide a complementary route to methods evaluation by leveraging numerical phantoms and high-fidelity simulations~\cite{abadi2020virtual}. By providing direct access to ground truth while permitting controlled variation of individual physical factors, VI studies enable systematic investigation of error mechanisms and sensitivity to modeling assumptions~\cite{abadi2020virtual}. Prior work has introduced advanced numerical phantoms for photoacoustic VI 
studies, including human breast phantoms \cite{Lou2017,Bao2021,park2023numerical,park2023stochastic,park2025virtual,akhlaghi2019multidomain}, head phantoms \cite{zhang2023photoacoustic,lyu2021photoacoustic,huang2025stochastic}, and prostate phantoms \cite{wu2024realistic}. However, realistic numerical phantoms tailored 
to multispectral DCE-PACT in 
preclinical small-animal imaging remain limited~\cite{cam2025numerical}. For this application, numerical mouse phantoms must provide temporally varying, wavelength-dependent optical properties and corresponding ground-truth CA concentration distributions, as well as account for respiratory motion.
This gap motivates the development of dedicated numerical mouse phantoms for controlled and physiologically meaningful VI studies.

The objective of this work is to develop a computational framework for generating numerical mouse phantoms
tailored for 
preclinical multispectral DCE-PACT. Building on a deformable whole-body mouse anatomical model with respiratory and cardiac motion~\cite{segars2004development,segars2009mcat}, the framework augments the anatomy with physiologically plausible vascular structure, heterogeneous tumor morphology, and time-varying CA 
kinetics modeled using the extended Tofts formulation\cite{tofts1999estimating}. Wavelength-dependent optical properties are then assigned 
to generate temporally resolved multispectral optical property maps based on spatiotemporally varying CA 
concentration distributions. The resulting numerical phantoms enable
reproducible VI studies for systematic evaluation of preclinical DCE-PACT methods under controlled variability. To demonstrate the utility of the proposed framework, an end-to-end VI case study is presented in which representative CA
concentration-estimation strategies are benchmarked, thereby illustrating how the proposed phantoms support quantitative method evaluation and investigation of fluence-related confounding effects in concentration estimation.

The remainder of the work is structured as follows. Section \ref{sec:background} provides background on the physical principles of multispectral DCE-PACT imaging and representative reconstruction approaches, followed by a review of existing numerical mouse anatomy phantoms with emphasis on whole-body models. Section \ref{sec:method} describes the methods used to construct the proposed numerical mouse phantoms, 
including dynamic anatomical phantom generation 
with embedded tumor and vasculature, CA
kinetics modeling using the extended Tofts model, functional-property assignment for chromophore content and oxygen saturation, and wavelength-dependent optical-property assignment. Section \ref{sec:usage} presents a VI 
case study demonstrating the utility of the framework, including photon transport and acoustic wave propagation simulations, acoustic reconstruction, 
CA
concentration estimation, and quantitative 
evaluation. Finally, Section \ref{sec:conclusions} discusses the utility, limitation, and possible extensions of the proposed phantom generation 
framework.

\section{Background}
\label{sec:background}
This section summarizes background material relevant to the  
proposed computational framework for generating numerical mouse phantoms 
for preclinical multispectral DCE-PACT. First, the physical principles of DCE-PACT imaging and representative reconstruction approaches are reviewed. Subsequently, numerical mouse anatomy phantoms are discussed 
with emphasis on the whole-body mouse phantom that serves as the anatomical basis of the framework developed in this work.

\subsection{Imaging Physics of Multispectral DCE-PACT and Reconstruction Methods}
\label{sec:imaging_physics}
In multispectral DCE-PACT, photoacoustic measurements are acquired at multiple optical wavelengths, typically in the near-infrared range (e.g., \qtyrange{700}{900}{nm})~\cite{jacques2013}, over a sequence of time points following administration of an exogenous CA.
The induced initial pressure at position $\mathbf{r}\in \mathbb{R}^3$, wavelength $\lambda$, and time $t$ can be expressed as
\begin{equation}
p_0(\mathbf{r},\lambda,t) = \Gamma(\mathbf{r})\,\Phi(\mathbf{r},\lambda,t;\mu_a)\,\mu_a(\mathbf{r},\lambda,t),
\label{eq:dce_forward}
\end{equation}
where $\Gamma(\mathbf{r})$ denotes the Gr\"uneisen parameter, $\mu_a(\mathbf{r},\lambda,t)$ is the optical absorption coefficient, and $\Phi(\mathbf{r},\lambda,t;\mu_a)$ is the optical fluence \cite{cox2012quantitative}. The dependence of $\Phi$ on $\mu_a$ arises because the fluence is governed by light transport through the global optical-property distribution. In DCE imaging, administration of a CA 
induces temporal changes in $\mu_a(\mathbf{r},\lambda,t)$ and thus perturbs $\Phi(\mathbf{r},\lambda,t;\mu_a)$ through optical absorption that depends on both wavelength and time.
Consequently, the measured temporal dynamics reflect both direct absorption changes and indirect fluence-mediated effects, and quantitative inference from dynamic multispectral measurements is complicated by this nonlinear absorption--fluence coupling \cite{cox2012quantitative,hupple2018light}.

In current practice, 
CA 
kinetics from multispectral PACT data are typically estimated in two steps 
by combining
acoustic reconstruction 
with a subsequent optical/spectral analysis.
In the first step, the measured acoustic data are inverted to  
estimate 
the induced initial pressure distribution $p_0(\mathbf{r},\lambda,t)$ at each wavelength and time point. In the second step, the reconstructed initial pressure images acquired across multiple wavelengths are used to estimate a CA-related quantity and its temporal evolution. Linear spectral unmixing is a widely used approach for this purpose.
This strategy is computationally simple and enables extraction of relative chromophore- or CA-associated signals, but it is intrinsically sensitive to wavelength-dependent fluence variations.
It can therefore produce biased, non-quantitative estimates when 
spectral coloring effects are significant \cite{cox2012quantitative, tzoumas2017spectral}. This limitation is particularly significant in DCE imaging, where quantitative PK 
analysis relies on concentration--time curves that are 
not confounded by spectrally dependent illumination \cite{hupple2018light}.

To address this issue, 
substantial
work has focused on mitigating fluence-related spectral distortion in multispectral photoacoustic signals through explicit or implicit fluence compensation. Model-based quantitative PACT incorporates photon transport modeling and estimates optical absorption, and in some cases scattering,
parameters that are consistent with the reconstructed initial pressure distribution \cite{cox2006two,bauer2011quantitative}. Hybrid strategies that combine photoacoustic measurements with diffuse optical tomography have also been proposed to provide independent estimates of fluence and thereby improve quantitative estimation of absorption \cite{bauer2011quantitative}. Other 
approaches seek to compensate for
fluence effects through simplified physical models or parameterizations of the fluence spectrum \cite{park2022normalization}. Examples include explicit fluence compensation
to improve quantitative chromophore estimation \cite{daoudi2012correcting}, and low-dimensional spectral modeling 
to reduce spectral coloring, such as eigenspectra-based multispectral optoacoustic tomography \cite{tzoumas2016eigenspectra}. Practical evaluation studies have also compared  
fluence-correction algorithms, including diffusion-based approximations and Monte Carlo approaches, 
highlighting trade-offs between model fidelity and computational cost \cite{zhou2020evaluation}.

In DCE applications, the objective is often to estimate PK 
descriptors by fitting compartmental models to CA-related time courses \cite{tofts1999estimating,sourbron2011scope}. Accordingly, multispectral DCE-PACT analysis pipelines must ultimately produce a temporal signal that is sufficiently related to CA 
concentration to support PK modeling. Strategies based on normalized temporal responses have been proposed to reduce sensitivity to unknown multiplicative fluence factors \cite{hupple2018light}.
For example, Hupple \emph{et al.} introduced a light-fluence-independent analysis framework for DCE multispectral optoacoustic tomography, in which a normalized dynamic response is constructed under the assumption that the dominant fluence-related scaling is approximately time-invariant. This approach permits
estimation of kinetic parameters without explicit recovery of absolute concentration \cite{hupple2018light}. Such approaches can be attractive in practice when absolute quantification is challenging, but their validity depends on whether 
the required assumptions hold in the presence of physiological motion, heterogeneous anatomy, and strongly time-varying absorption.

Collectively, prior work indicates that quantitative interpretation of multispectral DCE-PACT data requires careful consideration of (i) acoustic reconstruction accuracy, (ii) wavelength- and time-dependent optical fluence variations, 
and (iii) the kinetic models used to interpret dynamic signals. Realistic VI studies offer a means to systematically investigate these factors by supplying ground-truth CA distributions, temporally varying optical absorption, and PK parameters while enabling controlled variation of anatomy, motion, vascular morphology, and optical properties.

\subsection{Numerical Mouse Anatomy Phantoms}
\label{sec:bg_mouse_anatomy_phantoms}
Numerical mouse anatomy phantoms provide digital representations of murine anatomy that can be used to conduct VI 
studies with controlled and known underlying geometry and tissue composition.
For multispectral DCE-PACT, these phantoms must support whole-body anatomy with sufficient organ delineation to permit wavelength-dependent optical-property assignment and anatomically meaningful evaluation, while also accommodating physiological motion required for dynamic simulation studies.

Existing mouse anatomy phantoms differ primarily in how geometry is represented and whether physiological dynamics are modeled~\cite{dogdas2007digimouse, wang2015deformable, khmelinskii2011articulated}. A widely used deformable anatomical model is the MOBY phantom~\cite{segars2004development,Segars20074DMA,segars2009mcat}. MOBY represents mouse anatomy using non-uniform rational B-spline (NURBS) surfaces and incorporates parameterized respiratory and cardiac motion through deformation-based modeling, thereby enabling generation of time-resolved anatomical realizations with adjustable motion parameters \cite{segars2004development}.

Complementary to deformable models, voxelized anatomical atlases provide high-resolution static representations derived from imaging and/or histology. The Digimouse atlas, constructed from co-registered micro-CT and cryosection data with labeled organ segmentations, is a prominent example and has been widely used in simulation studies \cite{dogdas2007digimouse}. 
However, such atlases typically represent a single fixed anatomy and do not include intrinsic respiratory or cardiac motion.

For DCE-PACT applications, the ability to model physiological motion is particularly important because organ deformation and displacement can confound the interpretation of time-varying optical properties estimated from sequential dynamic acquisition protocols. 
In this regard, deformable phantoms such as MOBY provide a suitable anatomical foundation. Nevertheless, the MOBY phantom is not, by itself, sufficient for realistic multispectral DCE-PACT simulation. 
It does not include vascular networks or temporally varying optical-property maps that incorporate CA kinetics, nor does it support mechanisms for introducing tumor morphology and associated spatial heterogeneity in perfusion and permeability. 
These components are required to generate plausible ground-truth CA concentration fields and corresponding wavelength-dependent optical absorption distributions for quantitative evaluation of multispectral DCE-PACT methods.

\section{Methods}
\label{sec:method}

In this study, a DCE-PACT numerical mouse phantom is defined as a time-sequential collection of tissue label maps, referred to as dynamic anatomical phantoms, together with corresponding molecular and functional property distributions and wavelength-dependent optical property distributions. The time-sequential tissue label maps encode physiological motion, whereas the tissue property distributions are generated from these label maps while incorporating the effects of CA kinetics. The detailed modeling procedures are described in the following subsections.

\subsection{Dynamic Anatomical Phantom Construction}
\label{sec:dynamic_anatomical_phantom}

A dynamic whole-body mouse anatomical phantom was constructed by adopting the 4D digital MOBY phantom~\cite{segars2004development,Segars20074DMA,segars2009mcat} as a deformable anatomical scaffold, which was augmented with application-specific tumor and vasculature models.
Breathing motion was also adapted to mimic the effect of anesthesia. MOBY represents mouse anatomy using NURBS surfaces and provides parameterized cardiac and respiratory motion through deformation of organ and body-surface control points driven by user-defined time curves~\cite{segars2004development, Segars20074DMA}. The MOBY NURBS surfaces were discretized at each motion phase to yield a tissue label map
suitable for subsequent tissue property assignment.

\begin{figure}[tb]
\centering
\includegraphics[width=0.99\linewidth]{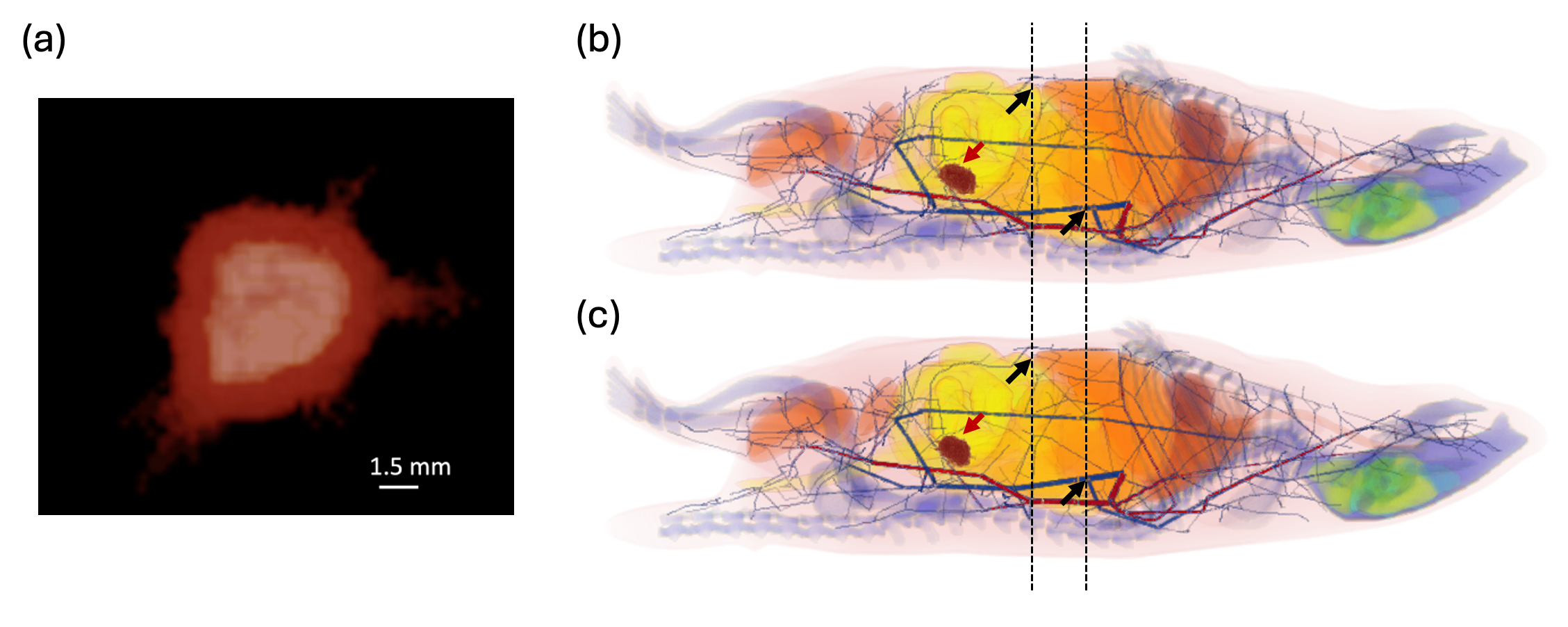}
\caption{Visualization of the constructed dynamic whole-body mouse phantom and embedded tumor model. (a) Example heterogeneous tumor realization with
an irregular viable-tumor region surrounding an internal necrotic core generated by morphological erosion. (b, c) 
Anatomical phantom with incorporated tumor and vasculature at
end-inspiration and end-expiration, respectively. The anatomy, tumor, and vascular structures were deformed consistently using the MOBY motion fields. Black arrows highlight representative regions where respiratory-induced displacement between the two states is most visually apparent. Red arrows indicate the location of the embedded tumor within the anatomical phantom.}
\label{fig:moby_tumor_visualization}
\end{figure}

To improve preclinical realism 
beyond the simple spherical tumor model available in MOBY, a heterogeneous spiculated tumor was embedded at
a user-defined soft-tissue location. Tumor boundaries were generated using tools adapted from the FDA VICTRE lesion-generation framework~\cite{badano2018evaluation}, which provides algorithmic procedures for generating realistic, irregular lesion morphologies for VI 
studies.
A necrotic core was then introduced by applying a morphological erosion operation to the viable-tumor mask using a spherical structuring element, yielding a two-compartment tumor with a viable rim and necrotic core~\cite{park2023stochastic}. In the example realization shown in Fig.~\ref{fig:moby_tumor_visualization} (a), the necrotic-core diameter was set to \qty{7.5}{mm} and the viable rim thickness was set to \qty{1.5}{mm}; these values were selected to produce a heterogeneous lesion of representative size in subcutaneous tumor models.

Because hemoglobin is a dominant endogenous absorber in the near-infrared wavelength range commonly used for PACT~\cite{jacques2013}, vascular structures contribute substantially to the optical absorption distribution and resulting photoacoustic signals. Therefore, explicit vascular structures were incorporated into the soft-tissue region. A synthetic arterial and venous tree pair was
generated using a constrained constructive optimization algorithm 
implemented in the open-source VItA library~\cite{talou2021adaptive}. In the present implementation, vessel diameters ranged from \qty{0.15}{mm} to \qty{0.45}{mm}. The lower bound
was set by the spatial discretization, whereas the upper bound
was selected to model larger conduit vessels.

Cardiac and respiratory motion were 
modeled by applying MOBY 
deformation fields to the anatomical phantom, including the newly incorporated tumor and vasculature, 
at each motion phase. In MOBY, respiratory deformation is controlled by time curves describing diaphragm displacement and anterior--posterior (AP) chest-wall expansion~\cite{segars2004development}. To better reflect anesthetized imaging conditions, the default MOBY respiratory motion drivers were modified by replacing the original AP and diaphragm control curves with custom NURBS control-point sets. Consistent with prior work that derived anesthetized-mouse respiratory motion from fluoroscopic measurements and implemented the resulting AP and diaphragm trajectories in MOBY, the overall rise-and-fall shape of the default waveform was preserved while the near-zero tail of the curve was extended to introduce a longer end-expiratory resting phase~\cite{van2017influence}. Here, the diaphragm trajectory was defined as the sign-inverted counterpart of the AP curve so that both respiratory drivers remained synchronized while moving in opposite coordinate directions.

The respiratory and cardiac periods were treated as user-selectable parameters. In this study, the respiratory period was set to \qty{2.5}{s} to reflect slowed breathing under deep anesthesia, and the cardiac period was set to \qty{0.175}{s}, 
corresponding to approximately \num{343} beats per minute, 
consistent with reported physiologic ranges for anesthetized mice~\cite{ewald2011monitoring, navarro2021mouse}. 
Representative visualizations of the resulting dynamic phantom at end-inspiration and end-expiration are
shown in Figs.~\ref{fig:moby_tumor_visualization} (b) and (c), 
illustrating anatomically coherent deformation of the body and incorporated tumor and vasculature.

This construction yields a time-resolved anatomical phantom that represents 
whole-body organ geometry, heterogeneous tumor morphology, explicit vascular structure, and physiological motion  
within a unified deformable model.
The parameter values above are intended as representative examples, and the framework remains flexible by allowing key modeling parameters, including tumor geometry, vascular caliber range, and motion periods, to be adjusted to emulate alternative experimental conditions and imaging protocols.

\subsection{Contrast-Agent Kinetics Modeling}
\label{sec:ca_kinetics_modeling}

To generate temporally varying CA concentration fields for DCE-PACT numerical mouse phantoms, CA kinetics were prescribed using
a compartmentalized PK
model comprising (i) an arterial input function (AIF) that describes
the CA concentration in plasma and (ii) tissue-specific
kinetic models that describe CA concentrations in specified tissue types, with tumor and liver modeled in this study.  
This construction yields ground-truth temporal CA concentration profiles that can be varied systematically by adjusting physiologically interpretable kinetic-model parameters, thereby enabling controlled use of the resulting numerical mouse phantoms in evaluations of DCE-PACT image reconstruction and concentration-estimation methods. 

The CA concentration in plasma, denoted $c_\mathrm{CA, p}(t)$ and expressed in molar units (\unit{M}), was modeled using a flexible empirical AIF that captures
slow bolus injection, tissue uptake, and clearance. Following the functional form employed in prior multispectral DCE-PACT studies \cite{hupple2018light}, $c_\mathrm{CA, p}(t)$ was defined as
\begin{equation}
c_\mathrm{CA, p}(t)=
\begin{cases}
0, & t \le T_0,
\\[4pt]
A\left(\frac{t-T_0}{60}\right)^{a_\mathrm{p}} e^{-B(t-T_0)}
+
C\left(1-e^{-D(t-T_0)}\right)e^{-E(t-T_0)},
& t > T_0,
\end{cases}
\label{eq:aif_model}
\end{equation}
where $t$ denotes dynamic time in seconds and $T_0$ is the bolus arrival time. The parameters $A$ and $C$ are scaling factors that control the relative magnitudes of the two kinetic components, whereas $a_\mathrm{p}$, $B$, $D$, and $E$ are kinetic parameters governing the temporal behavior of the CA concentration. 
Specifically, the parameters $A,a_\mathrm{p},B$ model the first-pass kinetics, consisting of a rapid rise as the bolus enters arterial circulation and a quick initial decay as the bolus disperses. The parameters $C,D,E$ model recirculation from venous return, gradual accumulation in slower compartments, and prolonged clearance through metabolism or excretion. 
Representative values are reported in Table \ref{tab:kinetics_param}.
In the VI
framework, these parameters were treated as user-selectable to generate
a range of physiologically plausible AIF realizations,
for example, by adopting representative parameter sets reported in prior DCE studies \cite{hupple2018light, jones2018linearization}.

CA kinetics within tumor tissue were modeled using the extended Tofts model (ETM), which is widely employed in DCE-MRI to characterize
vascular transport and permeability and which augments the original Tofts model by including an intravascular plasma contribution \cite{tofts1999estimating,sourbron2011scope}. Under the ETM, the CA concentration in the tumor, expressed in molar units (\unit{M}),
is given by
\begin{equation}
c_{\mathrm{CA, tumor}}(t)
=
v_\mathrm{p} c_\mathrm{CA, p}(t)
+
K_{\mathrm{trans}}
\left(
e^{(-k_{\mathrm{ep}} t)} *_t c_\mathrm{CA, p}(t)
\right),
\label{eq:etm_cont}
\end{equation}
where $*_t$ denotes temporal convolution, $v_\mathrm{p}$ is the plasma volume fraction, $K_{\mathrm{trans}}$ is the volume transfer constant, and $k_{\mathrm{ep}}$ is the efflux rate constant. In the numerical phantom implementation, the convolution term in Eq.~\eqref{eq:etm_cont} was evaluated on a uniform temporal grid $\{t_n\}_{n=0}^{N_t-1}$ with spacing $\Delta t$ using the discrete approximation
\begin{equation}
\left(e^{(-k_{\mathrm{ep}} t)} * c_\mathrm{CA, p}(t)\right)\Big|_{t=t_n}
\approx
\Delta t
\sum_{m=0}^{n}
e^{\left(-k_{\mathrm{ep}}(t_n-t_m)\right)}\,c_\mathrm{CA, p}(t_m).
\label{eq:etm_discrete}
\end{equation}
The ETM parameters $(v_\mathrm{p},K_{\mathrm{trans}},k_{\mathrm{ep}})$ were treated as adjustable quantities to generate
diverse kinetic behaviors and to support stress-testing of concentration estimation methods across physiologically meaningful variability \cite{sourbron2011scope}.

In addition to tumor CA  
kinetics, organ-dependent CA time courses can be prescribed to emulate distinct uptake and clearance behaviors associated with differences in vascular supply and tissue handling of the agent. In this framework, the liver was modeled separately because it is a primary site of
metabolic clearance
for commonly employed photoacoustic CAs, such as indocyanine green.
Accordingly, liver CA kinetics in the numerical mouse phantoms were represented using
a heuristic model designed
to capture hepatic CA accumulation
while retaining computational simplicity.
Let $t_{\mathrm{p,peak}}=\arg\max_t c_\mathrm{CA,p}(t)$ 
denote the AIF peak time. 
The
CA concentration in the liver, denoted $c_{\mathrm{CA, liver}}(t)$ and expressed in molar units (\unit{M}), was defined as
\begin{equation}
c_{\mathrm{CA, liver}}(t)
=
\begin{cases}
0, & t \le t_{\mathrm{p,peak}}+\Delta t_{\mathrm{liver}},\\[3pt]
\max\!\left(\beta\left(c_0 - c_\mathrm{CA, p}(t)\right),\,0\right), & t > t_{\mathrm{p,peak}}+\Delta t_{\mathrm{liver}},
\end{cases}
\label{eq:liver_heuristic}
\end{equation}
where $c_0$ represents the total injected CA level, $\beta>0$ is a scaling factor that accounts for the relative contribution of the blood and liver compartments, and $\Delta t_{\mathrm{liver}}\ge 0$ introduces an explicit delay relative to $t_{\mathrm{p,peak}}$.
This relationship was motivated by an approximate mass balance, assuming that changes in other compartments are negligible. Accordingly, as the plasma CA concentration decreases after the bolus peak, the corresponding liver concentration increases to approximately conserve the total CA content. The $\max(\cdot,0)$ operation ensures nonnegativity, while the delay term accounts for the delayed appearance of CA in the liver. Although this simplified relationship is not a mechanistic model of hepatic perfusion, it provides a controlled means to introduce organ-specific dynamics and to evaluate the sensitivity of DCE-PACT concentration estimation to heterogeneous kinetic behavior across anatomical regions.

Table~\ref{tab:kinetics_param} lists representative values of the CA kinetic modeling parameters used in this study. 
\begin{table}[t]
\centering
\caption{Representative values of the CA kinetic modeling parameters.}
\label{tab:kinetics_param}
\vspace{2pt}
\small
\begin{tabular}{lll}
\toprule
Parameter & Value & Eq. \\
\midrule
$T_0$               & \qty{0}{s}                    & \multirow{7}{*}{\eqref{eq:aif_model}} \\
$a_\mathrm{p}$      & \num{3}                       & \\
$A$                 & \qty{600d-6}{M/min}$^{a_\mathrm{p}}$ & \\
$B$                 & $4.34/60$~\unit{s^{-1}}       & \\
$C$                 & \qty{4.0d-6}{M}               & \\
$D$                 & $1/60$~\unit{s^{-1}}          & \\
$E$                 & $0.07/60$~\unit{s^{-1}}       & \\
\midrule
$K_{\mathrm{trans}}$ & $0.3/60$~\unit{s^{-1}}       & \multirow{2}{*}{\eqref{eq:etm_cont}} \\
$k_{\mathrm{ep}}$   & $0.75/60$~\unit{s^{-1}}       & \\
\midrule
$\beta$             & \num{2}                       & \multirow{3}{*}{\eqref{eq:liver_heuristic}} \\ 
$c_0$               & \qty{4.0d-6}{M}               & \\ 
$\Delta t_{\mathrm{liver}}$ & \qty{0}{s}            & \\ 
\bottomrule
\end{tabular}
\end{table}
By varying these parameters across realizations, the framework can emulate a range of uptake and wash-out behaviors relevant to preclinical DCE-PACT studies.

\subsection{Molecular and Functional Property Assignments}
\label{sec:functional_properties}

\begin{figure}[h!]
\centering
\includegraphics[width=0.99\linewidth]{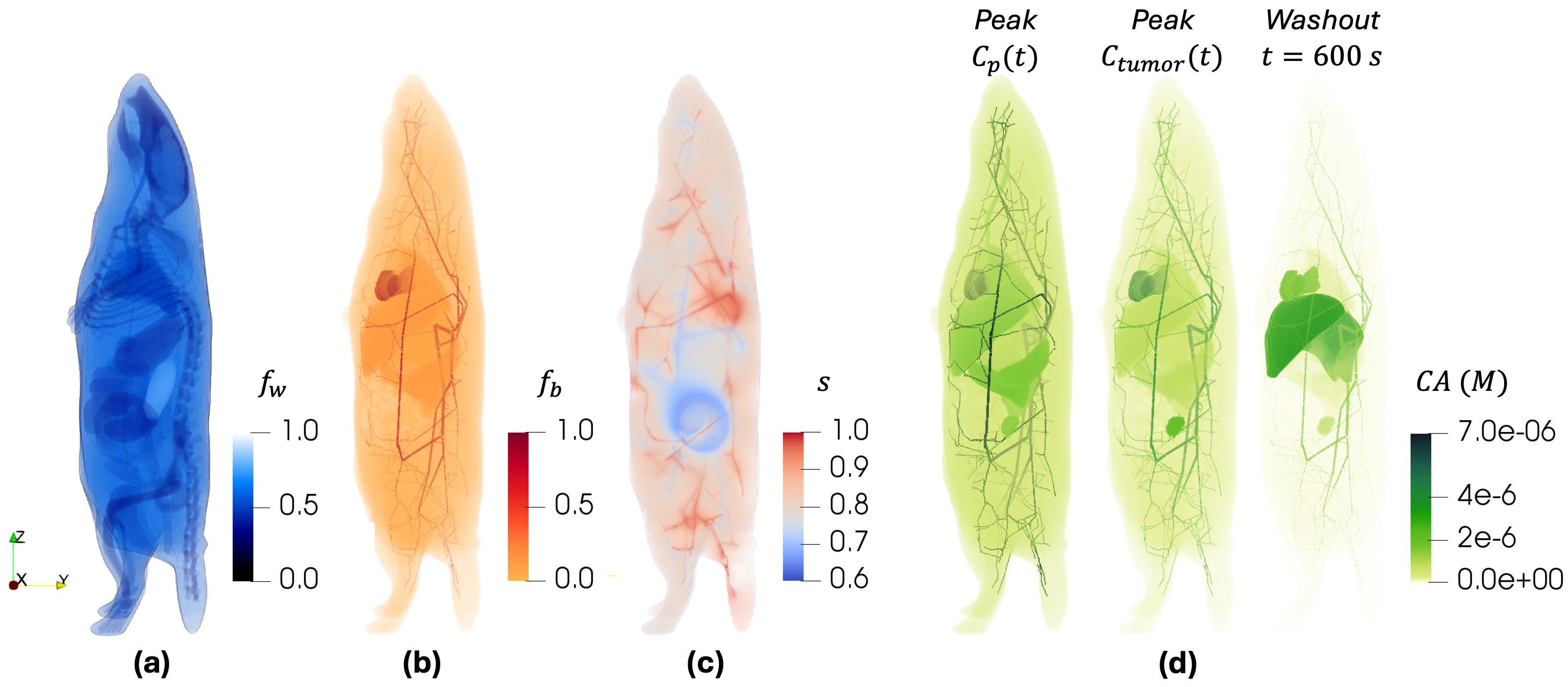}
\caption{Representative molecular and functional property distributions: (a) water volume fraction, $f_\mathrm{w}(\mathbf{r})$; (b) blood volume fraction, $f_\mathrm{b}(\mathbf{r})$; (c) oxygen saturation, $s(\mathbf{r})$; and (d) time-varying CA concentration, $c_\mathrm{CA}(\mathbf{r},t)$, at three time points: the plasma CA peak time, $t_\mathrm{p,peak}=43.2$~\unit{s}; the tumor CA peak time, $t_\mathrm{tumor,peak}=92$~\unit{s}; and a wash-out phase time point, $t_\mathrm{wash\mbox{-}out}=600$~\unit{s}.} 
\label{fig:functional_properties}
\end{figure}

Optical absorption in tissue arises primarily from endogenous chromophores and, when present, exogenous CAs. Its spatial distribution depends on the local composition and concentrations of these optical absorbers, which can be specified through \emph{molecular and functional} tissue parameters~\cite{jacques2013}. 
In the developed framework, oxygenated hemoglobin, deoxygenated hemoglobin, and water were considered endogenous chromophores because they are major optical absorbers in the near-infrared range (e.g., \qtyrange{700}{900}{nm}) commonly used in PACT~\cite{jacques2013}.
The concentrations of oxygenated hemoglobin, $c_{\mathrm{HbO_2}}(\mathbf{r})$ (\unit{M}), and deoxygenated hemoglobin, $c_{\mathrm{Hb}}(\mathbf{r})$ (\unit{M}), were determined by the tissue-dependent blood volume fraction, $f_{\mathrm{b}}(\mathbf{r})$ (unitless), and the spatially varying oxygen saturation field, $s(\mathbf{r})$ (unitless). For water, the tissue-dependent volume fraction, $f_{\mathrm{w}}(\mathbf{r})$ (unitless), was used directly in place of a molecular concentration. All of these quantities were assigned according to the dynamic anatomical phantom generated in Section~\ref{sec:dynamic_anatomical_phantom}. The exogenous CA concentration field, $c_\mathrm{CA}(\mathbf{r},t)$ (\unit{M}), was determined from both the dynamic anatomical phantom and the CA kinetic model described in Section~\ref{sec:ca_kinetics_modeling}. The resulting distributions of the endogenous chromophores and exogenous CA 
provide the basis for subsequent optical absorption calculations \cite{jacques2013, prahl1999tabulated, park2023stochastic}.
Table~\ref{tab:functional_props} summarizes the representative tissue-dependent values assigned to $f_{\mathrm{b}}(\mathbf{r})$ and $f_{\mathrm{w}}(\mathbf{r})$, selected based on literature-informed tissue compositions~\cite{jacques2013}. Figures~\ref{fig:functional_properties} (a) and (b) show the corresponding water and blood volume fraction maps. 
 
To obtain physiologically plausible blood oxygen saturation distributions with smooth transitions between tissue classes, 
an oxygen saturation field $s(\mathbf{r})\in[0,1]$ was computed
by solving a diffusion--reaction partial differential equation that approximates 
oxygen transport in biological tissue~\cite{Popel1989}. Following 
prior numerical phantom work \cite{park2023stochastic}, blood oxygen saturation was modeled 
as
\begin{equation}
-\mu \Delta s(\mathbf{r}) + \sum_{i\in\mathcal{T}_s} \chi_i(\mathbf{r})\big(s(\mathbf{r})-\bar{s}_i\big) = 0,
\label{eq:so2_pde}
\end{equation}
where $\mu>0$ is a smoothness parameter, $\mathcal{T}_s$ denotes a specified
set of tissue classes, $\chi_i(\mathbf{r})$ is the indicator function for tissue $i$, and $\bar{s}_i$ is the
corresponding target saturation. In this study, target oxygen saturation values were prescribed for
major vascular and tumor-associated tissue classes, with $\bar{s}_i=0.98$ for arteries, $\bar{s}_i=0.70$ for veins, $\bar{s}_i=0.37$ for viable tumor tissue, and $\bar{s}_i=0$ for the necrotic tumor core. This construction yields smoothly varying oxygenation distributions
while enforcing physiologically plausible average saturation levels in the tissue classes of interest. A representative oxygen saturation map is shown in Fig.~\ref{fig:functional_properties} (c). 

Given $f_{\mathrm{b}}(\mathbf{r})$ and $s(\mathbf{r})$, the concentrations of oxyhemoglobin and deoxyhemoglobin, expressed in molar units (\unit{M}), were defined as 
\begin{equation}
c_{\mathrm{HbO_2}}(\mathbf{r})
= f_{\mathrm{b}}(\mathbf{r})\,\frac{m_{\mathrm{Hb} }}{M_{\mathrm{Hb}}}\, s(\mathbf{r}), \quad
c_{\mathrm{Hb}}(\mathbf{r})
= f_{\mathrm{b}}(\mathbf{r})\,\frac{m_{\mathrm{Hb} }}{M_{\mathrm{Hb}}}\, \big(1-s(\mathbf{r})\big),
\label{eq:hb_split}
\end{equation}
where $M_{\mathrm{Hb}} = 64{,}500$~\unit{g/mol}
denotes the molecular weight of hemoglobin, and $m_{\mathrm{Hb}}=150$~\unit{g/L} denotes the total hemoglobin mass concentration in whole blood, consistent with reported values for mouse blood~\cite{raabe2011effects}.

Based on the CA kinetics models for plasma, tumor, and liver established in Section~\ref{sec:ca_kinetics_modeling}, the time-varying CA concentration distribution was defined by combining the temporal concentration profiles with spatial tissue indicators and blood volume fraction weighting as
\begin{equation}
c_{\mathrm{CA}}(\mathbf{r},t) = 
f_{\mathrm{b}}(\mathbf{r}) c_{\mathrm{CA, p}}(t) + \sum_{i \in \mathcal{T}_{\mathrm{CA}}}\chi_i(\mathbf{r}) c_{\mathrm{CA,}i}(t),
\label{eq:ca}
\end{equation}
where $\mathcal{T}_{\mathrm{CA}}$ denotes the set of tissue types of interest, i.e., tumor and liver; $c_{\mathrm{CA, p}}(t)$ denotes the plasma CA concentration defined in Eq. \eqref{eq:aif_model}; and $c_{\mathrm{CA,}i}(t)$ denotes the CA concentration in tissue type $i$, defined in Eq.~\eqref{eq:etm_cont} for tumor and Eq.~\eqref{eq:liver_heuristic} for liver. Figure~\ref{fig:functional_properties} (d) shows representative CA concentration maps at three time points: $t_\mathrm{p,peak} = 43.2~\unit{s}$, $t_\mathrm{tumor,peak} = 92~\unit{s}$, and $t_\mathrm{wash\mbox{-}out}=600$~\unit{s}, corresponding to the plasma CA peak time, tumor CA peak time, and a selected wash-out phase time point, respectively.

\vspace{4pt}
\begin{table}[t]
\centering
\caption{Representative volume fractions of blood and water.}
\label{tab:functional_props}
\vspace{2pt}
\small
\begin{tabular}{lll}
\toprule
Tissue & $f_{\mathrm{b}}$ & $f_{\mathrm{w}}$ \\
\midrule
Background (water) & \num{0}     & \num{1.00} \\
Body/skin          & \num{0.0337}& \num{0.60} \\
Viable tumor       & \num{0.0337}& \num{0.40} \\
Tumor core         & \num{0}     & \num{0.40} \\
Lung/airway        & \num{0.133} & \num{0.45} \\
Liver              & \num{0.154} & \num{0.60} \\
Kidney             & \num{0.119} & \num{0.80} \\
Artery             & \num{1.00}  & \num{0} \\
Vein               & \num{1.00}  & \num{0} \\
\bottomrule
\end{tabular}
\end{table}

\subsection{Optical Property Assignment}
\label{sec:optical_property_assignment}
The optical properties considered in this work consist of the spatially and temporally varying wavelength-dependent optical absorption coefficient $\mu_a(\mathbf{r},\lambda,t)$, the tissue-specific wavelength-dependent reduced scattering coefficient $\mu_s'(\mathbf{r},\lambda, t)$, and the tissue-specific scattering anisotropy factor $g(\mathbf{r}, t)$. The coefficients $\mu_a$ and $\mu_s'$ are expressed in $\mathrm{mm}^{-1}$, whereas $g$ is dimensionless. The wavelength dependence of $g(\mathbf{r},t)$ was neglected because scattering anisotropy varies only weakly over the near-infrared wavelength range considered in this study. Note that, whereas the temporal dynamics of $\mu_s'(\mathbf{r},\lambda, t)$ and $g(\mathbf{r},t)$ arise from respiratory and cardiac motion alone, the temporal dynamics of $\mu_a(\mathbf{r},\lambda,t)$ also depend on the CA kinetics.

\begin{figure}[h!]
\centering
\includegraphics[width=0.99\linewidth]{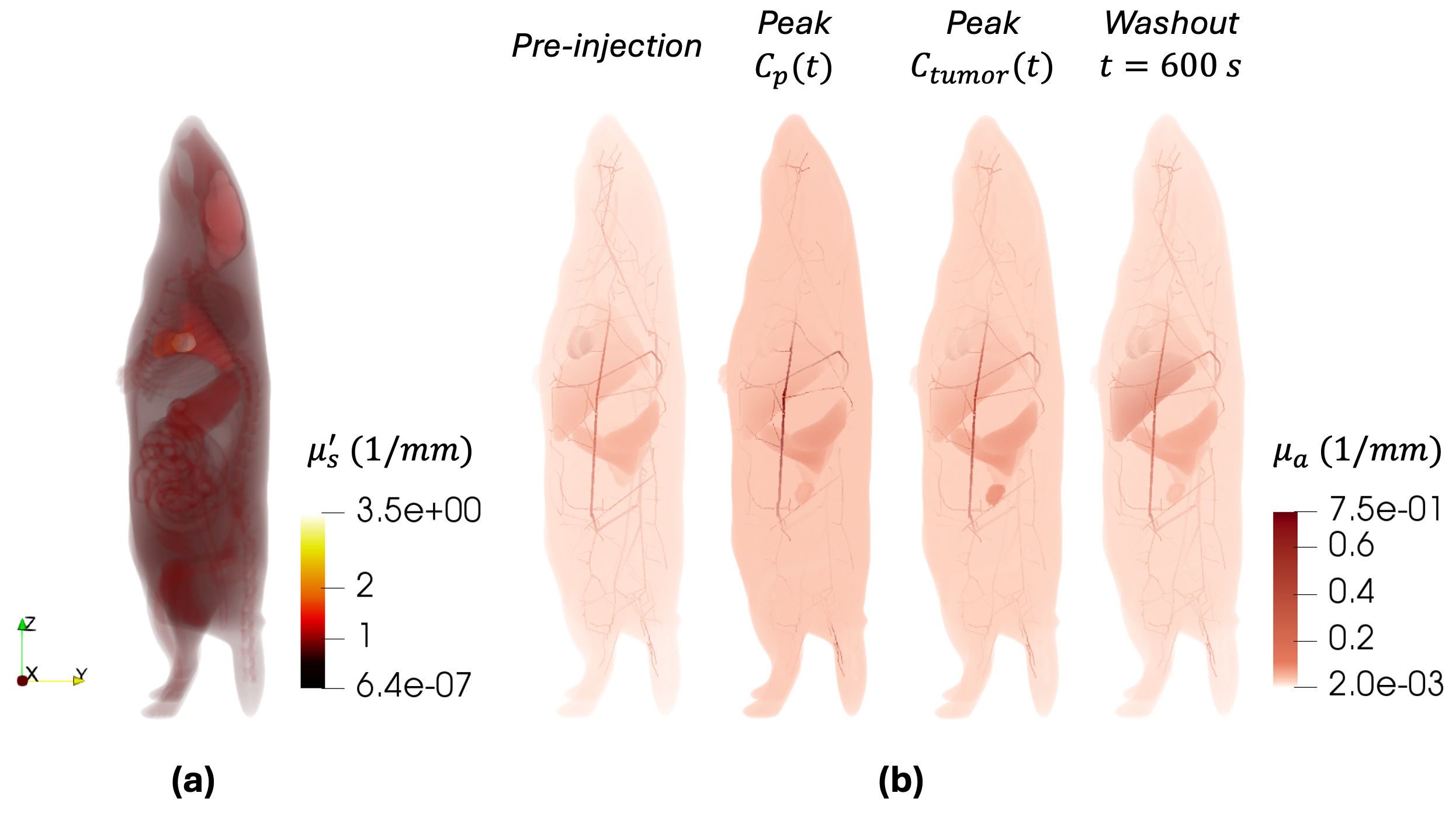}
\caption{Representative optical property distributions at $\lambda=800$~\unit{nm}: (a) reduced scattering coefficient, $\mu_s'(\mathbf{r},\lambda,t)$, and (b) time-varying optical absorption coefficient, $\mu_a(\mathbf{r},\lambda,t)$, at four time points: $t=0$~\unit{s} (pre-injection); the plasma CA peak time, $t_\mathrm{p,peak}=43.2$~\unit{s}; the tumor CA peak time, $t_\mathrm{tumor,peak}=92$~\unit{s}; and a wash-out phase time point, $t_\mathrm{wash\mbox{-}out}=600$~\unit{s}.} 
\label{fig:optical_properties}
\end{figure}

The optical absorption coefficient was modeled as the sum of contributions from the dominant endogenous chromophores and exogenous CA included in the phantom as follows~\cite{jacques2013}:
\begin{equation}
\mu_a(\mathbf{r},\lambda,t)
= \mathbf{e}^\top(\lambda) \mathbf{c}(\mathbf{r},t), \quad
\mathbf{e}(\lambda) = 
\begin{bmatrix}
\ln(10) \varepsilon_{\mathrm{HbO_2}}(\lambda) \\
\ln(10) \varepsilon_{\mathrm{Hb}}(\lambda) \\
\mu_{a,\mathrm{w}}(\lambda) \\
\ln(10) \varepsilon_{\mathrm{CA}}(\lambda)
\end{bmatrix}, \quad
\mathbf{c}(\mathbf{r},t) =
\begin{bmatrix}
c_{\mathrm{HbO_2}}(\mathbf{r}) \\
c_{\mathrm{Hb}}(\mathbf{r}) \\
f_{\mathrm{w}}(\mathbf{r}) \\
c_{\mathrm{CA}}(\mathbf{r},t) 
\end{bmatrix},
\label{eq:mua_assignment}
\end{equation}
where $\varepsilon_{\mathrm{HbO_2}}(\lambda)$, $\varepsilon_{\mathrm{Hb}}(\lambda)$, and $\varepsilon_{\mathrm{CA}}(\lambda)$ denote the molar extinction spectra of oxygenated hemoglobin, deoxygenated hemoglobin, and the CA,
respectively, expressed in \unit{mm^{-1}{M}^{-1}}; $\mu_{a,\mathrm{w}}(\lambda)$ denotes the optical absorption coefficient of pure water, expressed in \unit{mm^{-1}}; $c_{\mathrm{HbO_2}}(\mathbf{r})$ and $c_{\mathrm{Hb}}(\mathbf{r})$ denote
the hemoglobin concentration distributions
defined in Eq.~\eqref{eq:hb_split};
$f_{\mathrm{w}}(\mathbf{r})$ denotes
the water volume fraction distribution;
and $c_{\mathrm{CA}}(\mathbf{r},t)$ denotes
the time-varying CA
concentration field described in Eq.~\eqref{eq:ca}.
Under this formulation, temporal variation in $\mu_a(\mathbf{r},\lambda,t)$ arises from both
the evolving CA concentration distribution and the respiratory and cardiac motion.
Representative optical absorption maps at \qty{800}{nm} are shown
in Fig.~\ref{fig:optical_properties} (b) at four time points: $t=0$ (pre-injection), $t_{\mathrm{p,peak}}$, $t_{\mathrm{tumor,peak}}$, and $t_\mathrm{wash\mbox{-}out}$.

Scattering properties were assigned in a tissue-dependent manner through the commonly employed power-law model for reduced scattering~\cite{jacques2013},
\begin{equation}
\mu_s'(\mathbf{r},\lambda, t)
=
\mu_s'(\mathbf{r},\lambda_{\mathrm{ref}}, t)
\left(
\frac{\lambda_{\mathrm{ref}}}{\lambda}
\right)^{b(\mathbf{r})},
\label{eq:reduced_scattering_powerlaw}
\end{equation}
where $\mu_s'(\mathbf{r},\lambda_{\mathrm{ref}}, t)$ denotes 
the reduced scattering coefficient at the  
reference wavelength $\lambda_{\mathrm{ref}}$, and $b(\mathbf{r})$ denotes 
the scattering power-law exponent. In this study, $b=0.7$ was used for all tissue classes. Here, the temporal dependence of $\mu_s'(\mathbf{r},\lambda, t)$ arises solely from the phantom motion. The scattering coefficient, expressed in \unit{mm^{-1}}, was computed as
\begin{equation}
\mu_s(\mathbf{r},\lambda,t)
=
\frac{\mu_s'(\mathbf{r},\lambda,t)}{1-g(\mathbf{r},t)},
\label{eq:mus_from_musp}
\end{equation}
where $g(\mathbf{r},t)$ denotes the tissue-dependent anisotropy factor.
Each tissue class was therefore assigned
a triplet of
scattering parameters, $\{\mu_s'(\lambda_{\mathrm{ref}}), \lambda_{\mathrm{ref}}, g\}$, from which wavelength-dependent scattering distributions were
generated. Representative parameter values used in this study are summarized in Table~\ref{tab:scattering_props}. A representative map of the reduced scattering coefficient at $\lambda=800$~\unit{nm}
is shown in Fig.~\ref{fig:optical_properties} (a).

\begin{table}[t]
\centering
\caption{Representative scattering-model parameters.}
\label{tab:scattering_props}
\vspace{2pt}
\small
\begin{tabular}{llll}
\toprule
Tissue & $\mu_s'(\lambda_{\mathrm{ref}})$ (\unit{mm^{-1}}) & $\lambda_{\mathrm{ref}}$ (\unit{nm}) & $g$ (unitless) \\
\midrule
Background (water) & \num{7d-6}
& \num{700} & \num{0.90}  \\
Body/skin          & \num{0.66}              & \num{830} & \num{0.90}  \\
Viable tumor       & \num{0.75}              & \num{660} & \num{0.90}  \\
Tumor core         & \num{0.75}              & \num{660} & \num{0.90}  \\
Lung/airway        & \num{1.09}              & \num{850} & \num{0.90}  \\
Liver              & \num{0.56}              & \num{810} & \num{0.90}  \\
Kidney             & \num{0.54}              & \num{789} & \num{0.90}  \\
Artery             & \num{2.30}              & \num{500} & \num{0.975} \\
Vein               & \num{2.30}              & \num{500} & \num{0.975} \\
\bottomrule
\end{tabular}
\end{table}

\section{Example Virtual Imaging Case Study and Results}
\label{sec:usage}
To demonstrate the utility of the proposed computational framework for generating numerical mouse phantoms,
an example VI
study was conducted in which
an end-to-end multispectral DCE-PACT imaging pipeline was implemented. 
A representative DCE-PACT numerical mouse phantom was created on a Cartesian grid with isotropic voxel spacing of \qty{0.15}{mm} and a temporal resolution of \(\Delta t=0.1\)~\unit{s}.
As the present VI study was intended as an illustrative use case rather than a comprehensive dynamic study, the computationally intensive optical and acoustic simulations and image reconstruction pipeline were only performed at two representative time points: the plasma CA peak time, $t_\mathrm{p,peak}=43.2$~\unit{s}, and a time point during the wash-out phase, $t_\mathrm{wash\mbox{-}out}=600$~\unit{s}. Although periodic physiological motion was incorporated into the dynamic phantom, in this study the two time points were selected at the same motion phase, so that the effects of CA kinetics could be assessed without additional confounding from differences in motion phase.

At these selected time points, the study included 
optical illumination modeling, acoustic signal simulation, image reconstruction, and subsequent estimation of CA 
concentration from reconstructed multispectral images.
Throughout the VI study, the Gr\"uneisen parameter was assumed to be unity in both the photoacoustic forward model and the concentration estimation procedures. 
Representative CA 
concentration-estimation strategies 
with different levels of forward-model sophistication were
benchmarked against the known ground-truth CA 
distributions generated by the numerical phantoms.
The following subsections describe the simulation setup, estimation methods, evaluation strategy, and quantitative results.

\subsection{Photon Transport and Acoustic Wave Propagation Simulation}

\begin{figure}[h!]
\centering
\includegraphics[width=0.99\linewidth]{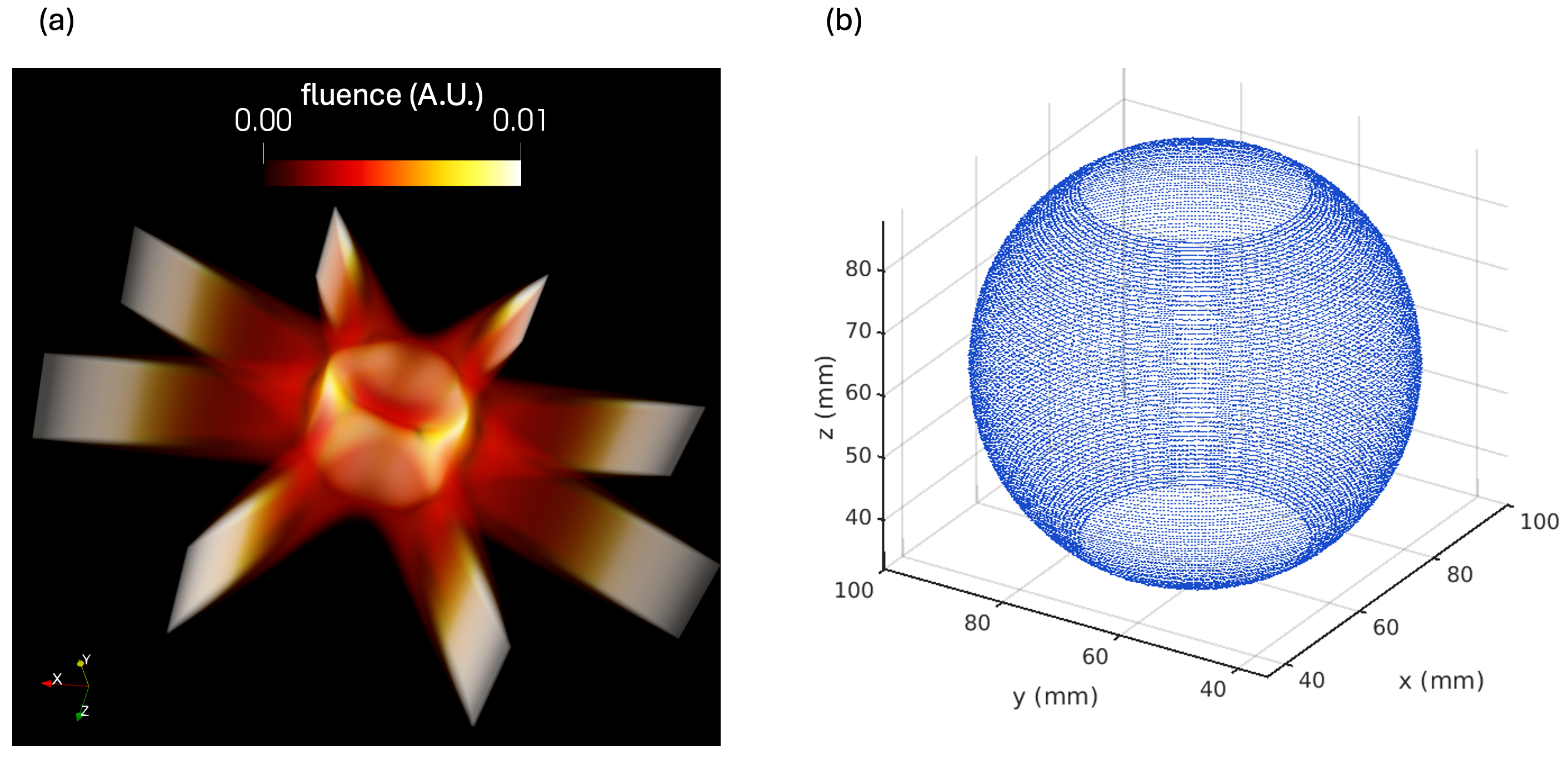}
\caption{Overview of the simulation configuration used in the example virtual imaging (VI) study. (a) Example optical fluence map obtained from MCX photon transport simulation using the eight-source Gaussian slit illumination configuration.
(b) Tomographic acoustic detection geometry used for acoustic signal simulation, modeled as a spherical segment measurement surface of radius \qty{65}{mm}, \num{96} detector locations per view, and \num{360} views.
}
\label{fig:optical_acoustic_simulation_geometry}
\end{figure}

For each of the two selected time points, multispectral optical simulations were performed using the corresponding dynamic numerical mouse phantom to generate wavelength-dependent
induced initial pressure maps.
Photon transport was modeled using the
GPU-accelerated Monte Carlo simulator
MCX~\cite{mcx1, mcx2}, at $N_\lambda=8$
excitation wavelengths, $\{730,750,770,780,790,800,820,$ $840\}$~\unit{nm}. Illumination was provided by eight Gaussian slit sources with a divergence of \ang{5} and a height of \qty{22.5}{mm},
centered along the axial direction and uniformly distributed on a circle of radius \qty{80}{mm} surrounding
the imaging volume, as shown in Fig.~\ref{fig:optical_acoustic_simulation_geometry} (a). The computational field of view was selected to encompass the tumor region and surrounding tissues and discretized into $1100 \times 1100 \times 350$ voxels, with a total of $N=423{,}500{,}000$ voxels. A spatially uniform refractive index of \num{1.37} was assumed. Following Ref.~\citenum{park2025virtual}, photon transport was simulated on a coarser grid with a voxel spacing of \qty{0.3}{mm} in each spatial dimension, and the resulting fluence maps were subsequently upsampled to the original resolution for computational efficiency. For each light source, \num{3d8} photons were simulated over a time window of \qty{50}{ns}. The induced initial pressure maps were then computed on the discretized grid from the corresponding optical absorption and fluence maps, following the continuous photoacoustic forward model in Eq.~\eqref{eq:dce_forward}.

Acoustic signals were generated
by simulating
wave propagation from
the induced initial pressure map for each wavelength and selected time point
under the assumption of an acoustically homogeneous and lossless medium,
using a GPU-accelerated implementation of an integral geometry forward model~\cite{wang2013accelerating}. A static tomographic detection geometry was employed, consisting of a spherical segment detection surface with a radius of \qty{65}{mm}, as shown in
Fig.~\ref{fig:optical_acoustic_simulation_geometry} (b).
Each tomographic view provided \ang{118} polar angular coverage with \num{96} detector elements, and full tomographic measurements were acquired over
\num{360} views. 
For each wavelength and selected time point, $N_\tau=2048$ acoustic time samples were acquired with a sampling interval of $\Delta\tau=0.032$~\unit{\micro\second}, corresponding to a total acquisition duration of $65.536$~\unit{\micro\second}. 
To emulate measurement noise, zero-mean
i.i.d. Gaussian noise was added
to the simulated acoustic data, with a standard deviation equal to 1\% of the maximum photoacoustic signal amplitude across all wavelengths and both selected time points~\cite{park2023stochastic, park2023numerical}.

\subsection{Acoustic Reconstruction and Contrast-Agent Concentration-Estimation Methods}

CA concentration was estimated
from the simulated acoustic measurements
using a two-step quantitative inversion framework consisting of acoustic image reconstruction followed by concentration estimation.

\subsubsection{Acoustic image reconstruction}
Let $\mathbf{g}(\lambda,t)\in\mathbb{R}^{N_d N_{\tau}}$ denote the measured acoustic data acquired at optical wavelength $\lambda$ and selected imaging time point
$t$, where $N_d=96 \times 360$ and $N_{\tau} = 2048$ denote, respectively, the total number of detector locations and acoustic time samples.  
Acoustic reconstruction was posed as estimation of the vectorized initial pressure map
$\mathbf{p}_0(\lambda,t)\in \mathbb{R}^{N}$. The estimate was obtained by solving the total-variation (TV) regularized least-squares problem
\begin{equation}
\hat{\mathbf{p}}_0(\lambda,t)
=
\arg\min_{\mathbf{p}_0(\lambda,t)}
\frac{1}{2}\left\|\mathbf{g}(\lambda,t)-\mathbf{H}\mathbf{p}_0(\lambda,t)\right\|_2^2
+
\alpha_{\mathrm{TV}}\left\|\mathbf{p}_0(\lambda,t)\right\|_{\mathrm{TV}},
\label{eq:tv_recon}
\end{equation}
where $\mathbf{H}\in \mathbb{R}^{N_d N_{\tau} \times N}$ denotes the discrete acoustic forward operator and $\alpha_{\mathrm{TV}}>0$ is a regularization parameter selected via grid search. Specifically, nine candidate values spanning $1\times10^{-9}$ to
$5\times10^{-7}$ were evaluated using a $1$-$2$-$5$ logarithmic
spacing for the reconstruction at $\lambda=730$~\unit{nm} and
$t=t_{\mathrm{p,peak}}$. The value yielding the lowest mean-squared
error (MSE) relative to the corresponding ground-truth initial
pressure distribution was selected, resulting in
$\alpha_{\mathrm{TV}}=1\times10^{-8}$. This value was subsequently
fixed and used for all other reconstructions. The minimization in Eq.~\eqref{eq:tv_recon} was performed using the FISTA-TV algorithm \cite{beck2009fast}.

\subsubsection{Concentration-estimation framework}
CA concentration maps were estimated from reconstructed multispectral initial pressure maps using wavelength-dependent spectral basis functions. Three representative concentration-estimation strategies with increasing levels of sophistication in modeling the optical fluence were evaluated: linear spectral unmixing~\cite{tzoumas2017spectral}, fluence-compensated spectral unmixing~\cite{park2022normalization, wang2023optical}, and an iterative physics-based method~\cite{cox2006two}. 

\paragraph{Linear spectral unmixing (SU)}
Linear spectral unmixing assumes spatially uniform and wavelength-independent optical fluence~\cite{tzoumas2017spectral}, yielding a linear relationship between the initial pressure and chromophore concentrations, i.e., $p_0(\mathbf{r}, \lambda, t) \propto \mathbf{e}^\top(\lambda) \mathbf{c}(\mathbf{r},t)$.
The reconstructed multispectral initial pressure maps were vectorized and stacked as $\hat{\mathbf{P}}_{0}(t)=\left[\hat{\mathbf{p}}_{0}(\lambda_1,t),\ldots,\hat{\mathbf{p}}_{0}(\lambda_{N_\lambda},t)\right]^\top \in \mathbb{R}^{N_{\lambda}\times N}$.
The constituent maps were collected as $\mathbf{C}(t)=\left[\mathbf{c}_{\mathrm{HbO_2}},\mathbf{c}_{\mathrm{Hb}},\mathbf{f}_{\mathrm{w}},\mathbf{c}_{\mathrm{CA}}(t)\right]^\top \in \mathbb{R}^{N_c\times N}$, where $N_c=4$ and each row has length $N$. The spectral basis matrix was defined as $\mathbf{E}=\left[\mathbf{e}(\lambda_1),\ldots,\mathbf{e}(\lambda_{N_\lambda})\right]^\top \in \mathbb{R}^{N_\lambda\times N_c}$. 
Under the fluence assumption, the constituent maps
were estimated by solving the following least-squares problem:

\begin{equation*}
\hat{\mathbf{C}}(t) = \arg \min_{\mathbf{C}(t)} \left\| \hat{\mathbf{P}}_{0}(t) - \mathbf{E}\mathbf{C}(t) \right\|_{F}^{2},
\label{eq:lsu}
\end{equation*}
where $\left\| \cdot \right\|_{F}$ denotes the Frobenius norm. The CA concentration estimate, $\hat{\mathbf{c}}_{\mathrm{CA}}(t)$, was obtained from the row of $\hat{\mathbf C}(t)$ corresponding to the CA spectral basis. 
In realistic imaging conditions, however,
optical fluence varies spatially and spectrally. As a result, the estimated concentrations may be biased
by spectral coloring effects and are not expected to be quantitatively accurate. Accordingly, SU was used
primarily as a baseline method for effect-size comparison rather than as a quantitative
estimator.
\paragraph{Fluence-compensated 
spectral unmixing (FC SU)}
To mitigate wavelength-dependent fluence effects, the reconstructed induced pressure maps were
normalized by use of estimated
optical fluence maps
\cite{park2022normalization, wang2023optical}. 
This approach assumed access to approximate, spatially averaged optical absorption and scattering coefficients representative of the pre-injection state.
Specifically, wavelength-specific two-region optical models were employed in which the mouse body was assigned spatially averaged absorption and scattering properties derived from the pre-injection state, while the surrounding water was assigned known wavelength-dependent optical properties. 
For each wavelength $\lambda$, an MCX~\cite{mcx1, mcx2} simulation
based on the
assumed spatially averaged optical properties was performed
to compute the estimated fluence map $\hat{\boldsymbol{\upphi}}(\lambda) \in \mathbb{R}^{N}$.
An optical absorption estimate $\hat{\boldsymbol{\upmu}}_{a}(\lambda,t) \in \mathbb{R}^{N}$ was then obtained by normalizing the reconstructed initial pressure map: 
\begin{equation*}
\hat{\boldsymbol{\upmu}}_{a}(\lambda,t) = \hat{\mathbf{p}}_{0}(\lambda,t) \oslash \left( \hat{\boldsymbol{\upphi}}(\lambda) + \delta \right),
\end{equation*}
where $\oslash$ denotes element-wise division, and $\delta>0$ is a small stabilization constant.
The constituent maps were subsequently estimated by solving
\begin{equation}
\hat{\mathbf{C}}(t) = \arg \min_{\mathbf{C}(t)} \left\| \hat{\mathbf{M}}_{a}(t) - \mathbf{E}\mathbf{C}(t) \right\|_{F}^{2},
\label{eq:lsu_fluence_norm}
\end{equation}
where $\hat{\mathbf{M}}_{a}(t) = \left[ \hat{\boldsymbol{\upmu}}_{a}(\lambda_1,t), \ldots, \hat{\boldsymbol{\upmu}}_{a}(\lambda_{N_\lambda},t) \right]^\top \in \mathbb{R}^{N_{\lambda} \times N}$ denotes the stacked multispectral absorption coefficient estimates. As in SU, the CA concentration estimate, $\hat{\mathbf{c}}_{\mathrm{CA}}(t)$, was obtained from the row of $\hat{\mathbf C}(t)$ corresponding to the CA spectral basis.

\paragraph{Iterative physics-based method}
An iterative physics-based approach~\cite{cox2006two} was evaluated to explicitly enforce consistency with the photoacoustic forward model while accounting for wavelength-dependent fluence effects. The method employed a fixed-point iteration that alternates between estimating the optical fluence and updating the optical absorption coefficient, while assuming spatially homogeneous and fixed optical scattering~\cite{cox2006two}.
Starting from an initial absorption coefficient estimate $\boldsymbol{\upmu}_{a}^{(0)}(\lambda,t) \in \mathbb{R}^{N}$, 
the optical fluence was computed as $\boldsymbol{\upphi}^{(k)}(\lambda,t) = \mathcal{F}(\boldsymbol{\upmu}_{a}^{(k)}(\lambda,t)) \in \mathbb{R}^{N}$, where $\mathcal{F}(\cdot)$ denotes 
the MCX forward light-transport operator.
Given this fluence estimate, the absorption coefficient was updated to enforce pointwise consistency with the reconstructed
initial pressure map:
\begin{equation*}
\boldsymbol{\upmu}_{a}^{(k+1)}(\lambda,t) = \hat{\mathbf{p}}_{0}(\lambda,t) \oslash \left(\boldsymbol{\upphi}^{(k)}(\lambda,t) + \delta \right).
\end{equation*}
Let
\begin{equation*}
\boldsymbol{\Phi}^{(k)}(t)=
\begin{bmatrix}
\boldsymbol{\upphi}^{(k)}(\lambda_1,t),
\ldots,
\boldsymbol{\upphi}^{(k)}(\lambda_{N_\lambda},t)
\end{bmatrix}^\top,
\qquad
\mathbf{M}_{a}^{(k)}(t)=
\begin{bmatrix}
\boldsymbol{\upmu}_{a}^{(k)}(\lambda_1,t),
\ldots,
\boldsymbol{\upmu}_{a}^{(k)}(\lambda_{N_\lambda},t)
\end{bmatrix}^\top,
\end{equation*}
both in $\mathbb{R}^{N_{\lambda} \times N}$, denote the wavelength-stacked fluence and absorption maps at iteration $k$. Iterations proceeded until the relative data misfit was
sufficiently small~\cite{cox2006two}:
\begin{equation*}
\frac{\left\| \hat{\mathbf{P}}_{0}(t)-\boldsymbol{\Phi}^{(k)}(t) \odot \mathbf{M}_{a}^{(k+1)}(t)\right\|_{F}}
{\left\| \hat{\mathbf{P}}_{0}(t) \right\|_{F}} \leq \epsilon,
\end{equation*}
where $\odot$ denotes the Hadamard product
and $\epsilon$ is the convergence tolerance. 
After convergence, the recovered absorption spectra were denoted by $\hat{\mathbf{M}}_{a}(t)$ and used as input to the least-squares spectral fitting problem in Eq.~\eqref{eq:lsu_fluence_norm} to estimate CA concentrations. 

\subsection{Evaluation Methodology}
\begin{figure}[h!]
\centering
\includegraphics[width=0.99\linewidth]{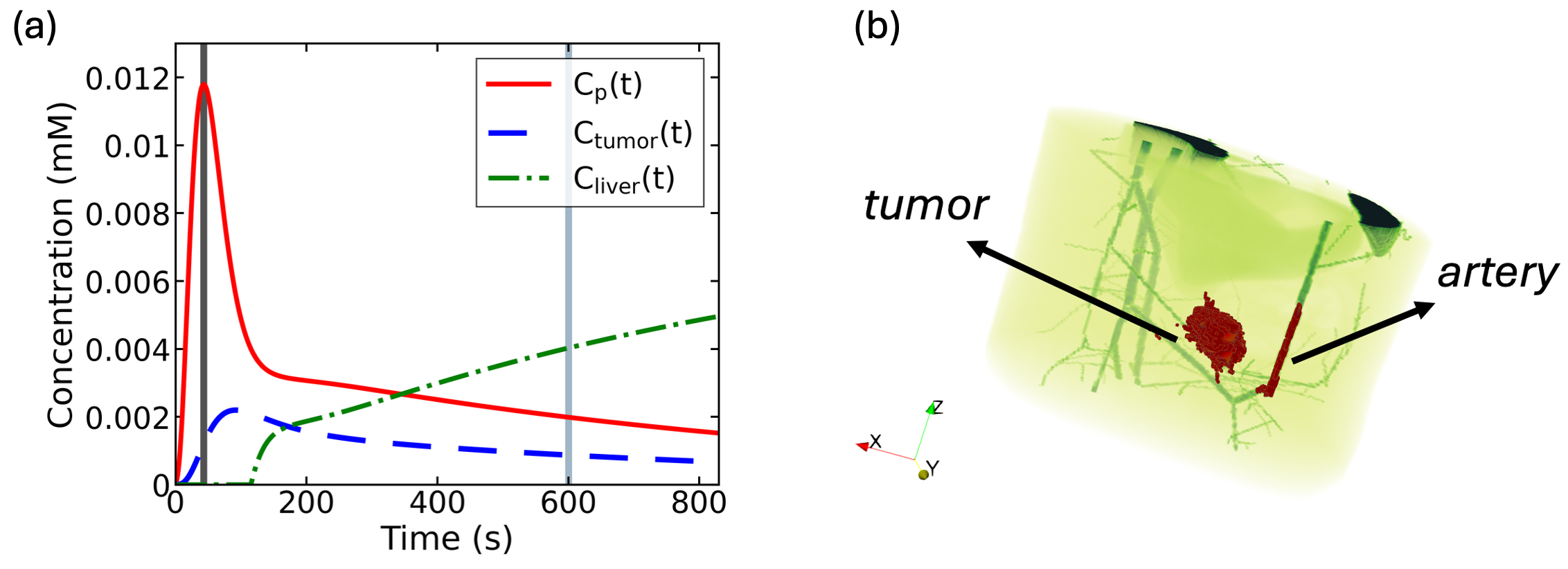}
\caption{Modeled CA kinetics and a CA concentration map overlaid with regions of interest (ROIs) used for quantitative evaluation: (a) CA 
concentration--time curves for
plasma (solid red line), tumor (solid blue line), and liver (green dash-dotted line); and (b) simulated CA concentration map at a representative wash-out time point $t_\mathrm{wash\mbox{-}out}=600$~\unit{s}.
In (a), the dark vertical line indicates the plasma CA peak time $t_\mathrm{p,peak}=43.2$~\unit{s},
and the light vertical line indicates $t_\mathrm{wash\mbox{-}out}$. In (b), the ROIs (dark red) included
the tumor and a proximal arterial region indicated by arrows.}
\label{fig:ca_timecourse_rois}
\end{figure}

The accuracy of CA concentration estimation
was quantitatively evaluated using known ground-truth CA concentration maps. To assess performance at distinct phases of CA kinetics,
two representative time points
were selected from the dynamic acquisition based on the CA concentration--time curve shown in Fig.~\ref{fig:ca_timecourse_rois} (a): (i) the plasma CA peak time, $t_\mathrm{p,peak}=43.2$~\unit{s}, corresponding to CA delivery and uptake, and (ii) a representative wash-out time point, $t_\mathrm{wash\mbox{-}out}=600$~\unit{s}, representing post-peak contrast clearance. 

\begin{figure*}[h!]
\centering
\includegraphics[width=\linewidth]{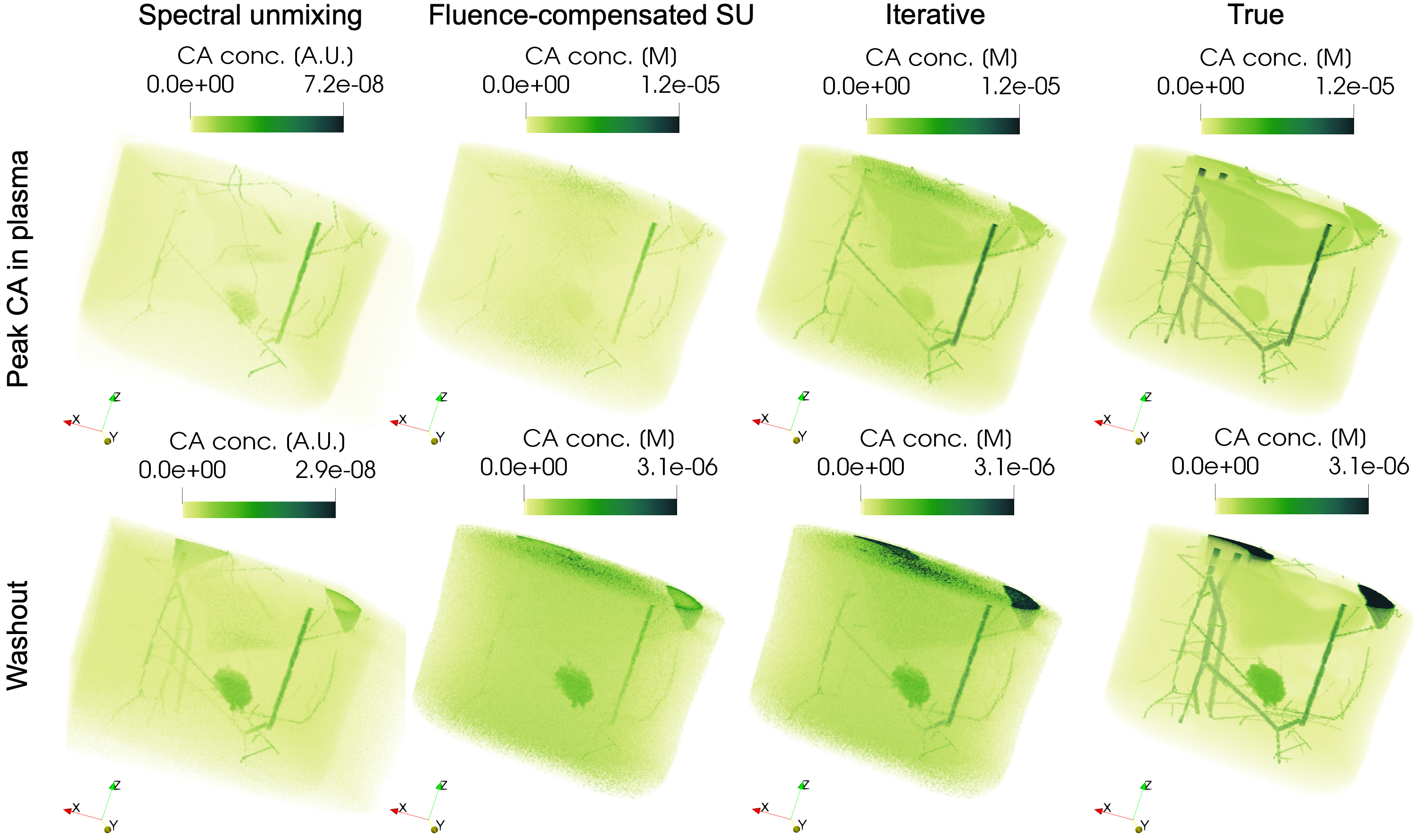}
\caption{Representative volumetric
CA
concentration estimates at two dynamic time points: the plasma CA peak time, $t_\mathrm{p,peak}=43.2$~\unit{s}
(top row), and a representative wash-out time, $t_\mathrm{wash\mbox{-}out}=600$~\unit{s} (bottom row). Columns show, from left to right, three
estimates obtained using linear spectral unmixing (SU; arbitrary units), fluence-compensated
spectral unmixing (FC SU; \unit{M}),
and the iterative physics-based method (\unit{M}), followed by the ground truth (\unit{M}).}
\label{fig:estimation_visual}
\end{figure*}

Quantitative evaluations were performed within a region restricted to voxels located within \qty{7.5}{mm} of the mouse skin surface, which included the tumor and nearby vascular structures. 
This depth restriction was motivated by
depth-dependent optical attenuation,
which leads to reduced photoacoustic signal strength and can diminish
quantitative reliability in deeper regions. Although accurate concentration estimation beyond this depth may be feasible under certain conditions, the selected region is relevant to preclinical applications involving
subcutaneous tumor models.

Two complementary metrics were used to evaluate CA concentration estimation performance: 
normalized root-mean-square error (nRMSE) and
Pearson correlation coefficient (PCC).
The nRMSE quantifies
concentration error relative to the ground-truth concentration magnitude at a given time point, whereas the PCC measures spatial agreement between estimated and ground-truth concentration maps
independent of global scaling. 
The nRMSE at time $t$ is defined as

\begin{equation*}
\mathrm{nRMSE}(t) = \frac{\|\hat{\mathbf{c}}_{\mathrm{CA}}(t) - {\mathbf{c}}_{\mathrm{CA}}(t)\|_{2}}{\|{\mathbf{c}}_{\mathrm{CA}}(t)\|_{2}},
\end{equation*}
and the PCC at time $t$ is defined as
\begin{equation*}
\mathrm{PCC}(t) = \frac{(\hat{\mathbf{c}}_{\mathrm{CA}}(t) - \overline{\hat{c}}_{\mathrm{CA}}(t))^\top(\mathbf{c}_{\mathrm{CA}}(t) - \overline{c}_{\mathrm{CA}}(t))}
{\|\hat{\mathbf{c}}_{\mathrm{CA}}(t) - \overline{\hat{c}}_{\mathrm{CA}}(t)\|_{2}\,
\|\mathbf{c}_{\mathrm{CA}}(t) - \overline{c}_{\mathrm{CA}}(t)\|_{2}},
\end{equation*}
where $\overline{\hat{c}}_{\mathrm{CA}}(t)$ and $\overline{c}_{\mathrm{CA}}(t)$
denote the spatial means of 
$\hat{\mathbf{c}}_{\mathrm{CA}}(t)$ and
$\mathbf{c}_{\mathrm{CA}}(t)$,
respectively. 

In addition to these
spatial metrics, tissue-level analysis was performed within
anatomically and physiologically relevant regions of interest (ROIs), marked in dark red in Fig.~\ref{fig:ca_timecourse_rois} (b).
Mean CA concentrations were computed within the tumor volume and a nearby arterial region.
The corresponding estimated and ground-truth mean ROI values are denoted by $\overline{\hat{c}}_{\mathrm{CA},i}(t)$ and $\overline{c}_{\mathrm{CA},i}(t)$, respectively, where $i\in\{\mathrm{tumor},\mathrm{artery}\}$. To assess whether CA uptake and wash-out dynamics were preserved, the percent change in the ground-truth mean ROI value between the plasma CA peak time and the wash-out time point was defined as
\begin{equation*}
\frac{\overline{c}_{\mathrm{CA},i}(t_{\mathrm{wash\mbox{-}out}}) - \overline{c}_{\mathrm{CA},i}(t_{\mathrm{p,peak}})}{\overline{c}_{\mathrm{CA},i}(t_{\mathrm{p,peak}})} \times 100\%.
\end{equation*}
The same calculation was applied to the estimated mean ROI value by replacing $\overline{c}_{\mathrm{CA},i}$ with $\overline{\hat{c}}_{\mathrm{CA},i}$.

\subsection{Results}
Representative volumetric CA concentration estimates at $t_\mathrm{p,peak}$ and $t_\mathrm{wash\mbox{-}out}$, together with the corresponding ground truth,
are shown in Fig.~\ref{fig:estimation_visual}. Quantitative performance in terms of nRMSE, PCC, and mean ROI concentrations at each time point is summarized in Table~\ref{tab:ca_results_stacked}. Because SU
does not compensate for wavelength-dependent fluence, its outputs are not quantitatively interpretable as molar concentrations.
Accordingly, the SU results are
interpreted primarily in terms of relative spatial agreement rather than absolute concentration accuracy.

\begin{table}[h!]
\centering
\begin{threeparttable}
\caption{Quantitative evaluation of CA concentration estimation at two representative dynamic time points.}
\label{tab:ca_results_stacked}
\vspace{2pt}
\centering
\small
\begin{tabular}{lcccc}
\toprule
\multicolumn{5}{c}{\makecell{Plasma CA peak time, $t_\mathrm{p,peak}=43.2$~\unit{s} \\
(
$\overline{c}_{\mathrm{CA,tumor}}=\mathbf{1.46\times10^{-6}}$~\unit{M}, 
$\overline{c}_{\mathrm{CA,artery}}=\mathbf{1.18\times10^{-5}}$~\unit{M} 
)}} \\
\midrule
Method & nRMSE & PCC & $\overline{\hat{c}}_{\mathrm{CA,tumor}}$ & $\overline{\hat{c}}_{\mathrm{CA,artery}}$ \\
\midrule
SU
& N/A             & 0.362          & $5.92\times10^{-9}$  & $2.32\times10^{-8}$\\
FC SU
& 0.043          & 0.272          & $8.26\times10^{-7}$~\unit{M}      & $2.09\times10^{-6}$~\unit{M} \\
Iterative
& \textbf{0.033} & \textbf{0.664} & $\mathbf{1.40\times10^{-6}}$~\unit{M} & $\mathbf{1.06\times10^{-5}}$~\unit{M} \\
\midrule
\multicolumn{5}{c}{\makecell{Representative wash-out time, $t_\mathrm{wash\mbox{-}out}=600$~\unit{s} \\
(
$\overline{c}_{\mathrm{CA,tumor}}=\mathbf{1.08\times10^{-6}}$~\unit{M}, 
$\overline{c}_{\mathrm{CA,artery}}=\mathbf{2.37\times10^{-6}}$~\unit{M} 
)}} \\
\midrule
Method & nRMSE & PCC & $\overline{\hat{c}}_{\mathrm{CA,tumor}}$ & $\overline{\hat{c}}_{\mathrm{CA,artery}}$ \\
\midrule
SU
& N/A             & 0.428          & $6.09\times10^{-9}$  & $1.44\times10^{-8}$\\
FC SU
& 0.098          & 0.424          & $8.87\times10^{-7}$~\unit{M}      & $7.85\times10^{-7}$~\unit{M} \\
Iterative
& \textbf{0.080} & \textbf{0.760} & $\mathbf{1.07\times10^{-6}}$~\unit{M} & $\mathbf{2.17\times10^{-6}}$~\unit{M} \\
\bottomrule
\end{tabular}
\begin{tablenotes}
\footnotesize
\item N/A: not applicable. For SU, nRMSE is not reported because the estimated values are not expressed on the same concentration scale as the ground truth; mean ROI values are reported in arbitrary units. Boldface values indicate the best-performing result for each metric.
\end{tablenotes}
\end{threeparttable}
\end{table}

Several trends are apparent. First, across both time points, the iterative physics-based method
provides the most accurate concentration estimates across all metrics,
indicating that explicitly accounting for wavelength-dependent fluence is important for recovering quantitative CA concentration maps. Second, although FC SU
yields molar-valued estimates and improves quantitative interpretability relative to SU, residual bias remains. This behavior is consistent with the fact that simplified fluence normalization can only partially mitigate spectral corruption caused by heterogeneous, wavelength-dependent optical attenuation.

The qualitative comparisons in Fig.~\ref{fig:estimation_visual} corroborate these findings.
SU captures relative contrast but exhibits
arbitrary scaling and noticeable spatial inaccuracies. FC SU partially restores concentration magnitude but
still
attenuates vascular contrast and introduces local concentration errors, suggesting that fluence normalization alone is insufficient to fully recover quantitative CA concentration maps.
By contrast, the iterative physics-based method more faithfully recovers vascular structures
and
concentration distributions, yielding maps
that are visually consistent with the ground truth.

\begin{table}[h!]
\centering
\caption{Percent change in the mean ROI values between the plasma CA peak time and the wash-out time point.}
\label{tab:ca_pct_change}

\begin{tabular}{lcc}
\toprule
 & Tumor ($\%$)
 & Artery ($\%$) \\
\midrule
Ground truth & $-26.03 $ & $-79.92 $ \\
\midrule
SU        & $+2.87 $  & $-37.93 $ \\
FC SU     & $+7.38 $  & $-62.44 $ \\
Iterative & $\mathbf{-23.57 }$ & $\mathbf{-79.53 }$ \\
\bottomrule
\end{tabular}
\end{table}

Table~\ref{tab:ca_pct_change} reports the percent change in the mean ROI values between the plasma CA peak time and the wash-out time point. Because
this metric is invariant to global scaling, it provides a complementary evaluation of temporal
fidelity for all methods, including SU. The
iterative physics-based method most accurately
reproduces both the direction and magnitude of the ground-truth changes in the tumor and arterial
regions, whereas SU and FC SU exhibit
biased estimates of the temporal changes
in at least one region.
Collectively, these findings reinforce that, in multispectral DCE-PACT, both quantitative concentration recovery and the characterization
of physiologically meaningful temporal trends are sensitive to the fluence distribution. The results further suggest
that physics-based fluence modeling provides a substantial performance advantage, consistent with prior observations in quantitative multispectral photoacoustic imaging \cite{cox2012quantitative, hupple2018light}.

\section{Discussion and Conclusion}
\label{sec:conclusions}
This study introduced a computational framework for generating numerical mouse phantoms
for VI
studies of preclinical multispectral DCE-PACT. The
main contribution is a simulation-ready pipeline that integrates deformable whole-body anatomy, physiological motion, explicit vasculature, heterogeneous tumor morphology, and time-varying CA concentration
distributions into multispectral optical-property maps. Accordingly, the framework provides
a controllable computational testbed for evaluating multispectral DCE-PACT methods under application-relevant variability \cite{hacker2024tutorial, manohar2024super}.

Such VI
imaging studies are valuable in photoacoustic imaging because rigorous validation is difficult to perform \emph{in vivo}, where the true optical, physiological, and PK
quantities are generally unknown. Moreover, simplified physical phantoms do not readily reproduce the coupled anatomical, optical, and temporal complexities encountered in live small-animal studies \cite{hacker2024tutorial, manohar2024super}.
In this
context, the example VI
study should be interpreted primarily as an illustration of the framework's utility. Specifically,
it demonstrates how the developed phantoms can be used to investigate the influence of wavelength-dependent fluence variations on the recovery of CA
distributions and temporal uptake behavior in multispectral DCE-PACT, which is relevant for downstream quantitative and PK
analyses \cite{tarvainen2024quantitative, hupple2018light, xiao2018applying}.

An important feature of the developed framework is its modularity. In the present study, the emphasis was placed on anatomical modeling and optical-property assignment, since these components are foundational for generating realistic multispectral photoacoustic data and for studying light-transport-related confounding.
Nevertheless, the framework can accommodate more comprehensive multiphysics modeling.
For example, future studies can extend the framework by incorporating 
heterogeneous acoustic properties, including spatially varying speed of sound, density, and attenuation,
from literature values according to the target application. Such an extension would be
important because acoustic heterogeneity is a recognized source of aberration, resolution loss, and model mismatch in photoacoustic image reconstruction \cite{huang2013full, wang2020combating}.
Beyond acoustic modeling, the modular design can also support 
organ-specific pharmacokinetic descriptions, task-specific VI
protocols \cite{hacker2024tutorial, manohar2024super}, and stochastic generation of virtual populations of dynamic numerical mouse phantoms.

\subsection*{Code, Data, and Materials Availability}

The codebase used to generate the numerical mouse phantoms described in this study is publicly available at \url{https://github.com/precise-lab/dce-moby4d}. This repository includes scripts for constructing the dynamic numerical phantoms, assigning functional and optical properties, and generating contrast-agent concentration distributions. Additional data required to reproduce specific results presented in this study are available from the corresponding author upon reasonable request.

\subsection*{Disclosures}
The authors declare that there are no financial interests, commercial affiliations, or other potential conflicts of interest that could have influenced the objectivity
 of this research or the writing of this paper.

\paragraph{Declaration of use of Generative AI}
A large-language model (ChatGPT) was used to proofread, check grammar, and improve the flow of selected passages of this work. The authors independently reviewed, verified, and take full responsibility for every edited sentence.

\subsection*{Acknowledgments}
This work was supported in part by the National Institutes of Health (United States) under grants EB034261, EB031585, EB031772, and EB028744. The authors acknowledge the Texas Advanced Computing Center (TACC) at The University of Texas at Austin for providing computational resources on the Vista system that have contributed to the research results reported within this paper (https://www.tacc.utexas.edu). Vista is supported by the U.S. National Science Foundation as a supplement to Frontera: Computation for the Endless Frontier (Award No. 1818253). The virtual imaging study reported in this work used the Delta system at the National Center for Supercomputing Applications through allocation MDE230007 from the Advanced Cyberinfrastructure Coordination Ecosystem: Services \& Support (ACCESS) program, which is supported by U.S. National Science Foundation grants \#2138259, \#2138286, \#2138307, \#2137603, and \#2138296. 

\bibliography{report}
\bibliographystyle{spiejour}

\listoffigures
\listoftables

\end{document}